\documentclass{aa}  

\usepackage{graphicx}
\usepackage{txfonts}

\usepackage{color, xcolor}
\usepackage[breaklinks=true]{hyperref}
\usepackage{breakurl}
\usepackage{ulem} 

\usepackage{natbib,twoopt}

\usepackage{flushend}

\usepackage{caption}  

\usepackage{array}   

\hypersetup{
  colorlinks = true,
  linkcolor=blue,   
  citecolor=blue,   
  urlcolor=blue,    
  pagebackref=true,
  implicit=false,
  bookmarks=true,
  bookmarksopen=true,
  pdfdisplaydoctitle=true
}

\bibpunct{(}{)}{;}{a}{}{,}    
\makeatletter
 \newcommandtwoopt{\citeads}[3][][]{\href{http://adsabs.harvard.edu/abs/#3}%
   {\def\hyper@linkstart##1##2{}%
    \let\hyper@linkend\@empty\citealp[#1][#2]{#3}}}    
 \newcommandtwoopt{\citepads}[3][][]{\href{http://adsabs.harvard.edu/abs/#3}%
   {\def\hyper@linkstart##1##2{}%
    \let\hyper@linkend\@empty\citep[#1][#2]{#3}}}      
 \newcommandtwoopt{\citetads}[3][][]{\href{http://adsabs.harvard.edu/abs/#3}%
   {\def\hyper@linkstart##1##2{}%
    \let\hyper@linkend\@empty\citet[#1][#2]{#3}}}      
 \newcommandtwoopt{\citeyearads}[3][][]%
   {\href{http://adsabs.harvard.edu/abs/#3}%
   {\def\hyper@linkstart##1##2{}%
    \let\hyper@linkend\@empty\citeyear[#1][#2]{#3}}}   
\makeatother

\graphicspath{ {./Figs/} }

\begin{document}

\title{Measuring and characterizing the line profile of HARPS \\ with a laser frequency comb} 


\author{F. Zhao \inst{1}
         \and
          G. Lo Curto \inst{2} 
          \and
          L.Pasquini \inst{2}
          \and
          J. I. Gonz\'{a}lez Hern\'{a}ndez \inst{4,5}
          \and
          J. R. De Medeiros \inst{6}
          \and
          B. L. Canto Martins \inst{6}
          \and
          I. C. Le\~{a}o \inst{6}
          \and
          R. Rebolo\inst{4,5,7}
          \and
          A. Su\'{a}rez Mascare\~{n}o\inst{4,5}
          \and
          M. Esposito\inst{4,8}
          \and
          A. Manescau \inst{2}
          \and
          T. Steinmetz \inst{3,9}
          \and
          T. Udem \inst{9}
          \and  \\
          R. Probst \inst{3}
          \and
          R. Holzwarth \inst{3,9}
          \and
          G. Zhao \inst{1}
          }

\institute{CAS Key Laboratory of Optical Astronomy, National Astronomical Observatories, Chinese Academy of Sciences, \\ Beijing 100101, China
         \and
             European Southern Observatory, Karl-Schwarzschild-Str. 2, 85748 Garching, Germany
         \and
             Menlo Systems GmbH, Am Klopferspitz 19a, 82152 Martinsried, Germany
         \and
             Instituto de Astrof{\'\i}sica de Canarias (IAC), E-38205 La Laguna, Tenerife, Spain
         \and
             Universidad de La Laguna, Dpto. Astrof\'{i}sica, E-38206 La Laguna, Tenerife, Spain
         \and
             Departamento de F\'{i}sica, Universidade Federal do Rio Grande do Norte, Campus Universit\'{a}rio, Natal, RN, 59078-970, Brazil
         \and
             Consejo Superior de Investigaciones Cient\'{i}ficas, Spain
         \and
             INAF -- Osservatorio Astronomico di Capodimonte, Salita Moiariello 16, I-80131, Napoli, Italy
         \and
             Max-Planck-Institut f\"ur Quantenoptik, Hans Kopfermann Str. 1, 85748 Garching, Germany \\
             }

\date{Submitted Jul. 7th, 2020}

\abstract
   {}
   {
We study the 2D spectral line profile of HARPS (High Accuracy Radial Velocity Planet Searcher), measuring its variation with position across the detector and with changing line intensity. 
The characterization of the line profile and its variations are important for achieving the precision of the wavelength scales of $10^{-10}$ or 3.0 $cm \ s^{-1}$ necessary to detect Earth-twins in the habitable zone around solar-like stars. 
}
   {We used a laser frequency comb (LFC) with unresolved and unblended lines to probe the instrument line profile.
We injected the LFC light -- attenuated by various neutral density filters -- into both the object and the reference fibres of HARPS, and we studied the variations of the line profiles with the line intensities. 
We applied moment analysis to measure the line positions, widths, and skewness as well as to characterize the line profile distortions  induced by the spectrograph and detectors.
Based on this, we established a model to correct for point spread function distortions by tracking the beam profiles in both fibres.}
   {We demonstrate that the line profile varies with the position on the detector and as a function of line intensities.
This is consistent with a charge transfer inefficiency (CTI) effect on the HARPS detector. 
The estimate of the line position depends critically on the line profile, and therefore a change in the line amplitude effectively changes the measured position of the lines, affecting the stability of the wavelength scale of the instrument.
We deduce and apply the correcting functions to re-calibrate and mitigate this effect, reducing it to a level consistent with photon noise.}
   {}

\titlerunning{Measuring and characterizing the line profile of HARPS} 
\maketitle
%

\section{Introduction}   \label{sec:intr}

High-precision spectroscopy is one of the most successful methods for detecting exoplanets. Over the past two and a half decades, developments in radial velocity (RV hereafter) measurement techniques have triggered a multitude of discoveries. Several milestones are particularly remarkable, such as the first evidence of giant planets orbiting solar-type stars \citep{1995Natur.378..355M}, the detection of a population of Neptunes and super-Earths  \citep{2011A&A...528A.112L}, and the discovery of Earth-mass planets in habitable zones \citep{2016Natur.536..437A, 2017Natur.542..456G}.
These discoveries came in parallel with efforts to improve the precision of Doppler measurements, which is required to detect low-mass exoplanets \citep{2014Natur.513..328M}.
Thanks to the development of extreme precision spectrographs such as HARPS (High Accuracy Radial Velocity Planet Searcher), Doppler measurement precision has improved to better than  1 $ \rm m \ s^{-1} $\citep{2002Msngr.110....9P, 2003Msngr.114...20M, 2006Natur.441..305L}.
Doppler detection of an exoplanet is obtained by measuring the amplitude of the tiny wobble induced on the host star. For an Earth analogue orbiting a solar-type star, this detection requires an RV precision of $\approx$3 $ \rm cm \ s^{-1} $ (or a one part in $10^{10}$ level).


The state-of-the-art HARPS spectrograph, mounted at the European Southern Observatory 3.6 m telescope in La Silla, is a high-resolution cross-dispersed echelle spectrograph that is optimized for exoplanet searches \citep{2003Msngr.114...20M}. This fibre-fed spectrograph has a median spectral resolution of 115,000 and records 72 echelle orders, covering the spectral range of 3775-6905\AA. It is located in a vacuum vessel under strict pressure and temperature control to maintain its thermo-mechanical stability over long timescales. Thanks to the high stability of the spectrograph and the detector mosaic -- two 2k$\times$4k EEV
charge-coupled devices (CCDs)
-- as well as the simultaneous calibration method \citep{1996A&AS..119..373B},
a single-measurement RV precision of 60 $ \rm cm \ s^{-1} $ has been achieved.

The RV precision of HARPS is primarily limited by i)\ photon noise, ii) the\ stability of the light injected into the spectrograph, and iii)\ wavelength calibration \citep{2010SPIE.7735E..33L}.
This work demonstrates that detector effects (such as charge transfer inefficiency) also play a significant role since they can modify the shape of the point spread function (PSF) and therefore alter wavelength calibration line positions between exposures.

Iodine absorption cells and Thorium-Argon hollow-cathode  (ThAr) lamps have been widely used to establish large sets of spectral lines for wavelength reference and calibration. However, their limitations are noticeable when we aim for better than metre per second
RV precision.
Line blending, non-uniform line spacing, and non-uniform line intensity limit the global precision of ThAr lamps to, at best, $10^{-9}$ for 10,000 lines, or uncertainties of up to 100 $ \rm m \ s^{-1}$ on individual lines \citep{1983ats..book.....P}.
Therefore, for centimetre per second
RV precision, a novel wavelength calibration source is essential.

An ideal wavelength calibrator \citep{2007MNRAS.380..839M, 2008eic..work..375P, 2012Msngr.149....2L} should possess the following three main characteristics: i) as many unresolved and unblended lines as possible; ii) uniform line intensity and line spacing; and iii) stability at the $10^{-11}$ level on both short and long timescales.
Today, the development of laser frequency combs (LFCs) \citep{2000PhRvL..84.3232R, 2002fqml.conf..253U} moves us closer towards optimum wavelength calibration.
Laser frequency combs for astronomy are extremely precise  and stable wavelength calibration sources that can generate thousands of uniformly spaced lines (also called modes) in a frequency space within the wavelength domain covered by a modern cross-dispersed astronomical spectrograph. The line frequencies can be locked to an atomic clock, from which they inherit their stability at short and long timescales \citep{2010MNRAS.405L..16W}.
Therefore, precision and accuracy of up to one part in $10^{11}$ are, in principle, achievable.

The mode frequencies of a typical LFC spectrum can be expressed in the form of
\begin{equation}
\mathrm{f=f_{ceo}+n \cdot f_{rep}}
 \label{e1}
, \end{equation}
where $\mathrm{f_{ceo}}$ is the carrier-envelope offset frequency \citep{2002fqml.conf..253U} and $\mathrm{f_{rep}}$ stands for the repetition frequency, which defines the frequency distance between two modes. Both parameters can be synchronized to an atomic clock and can be defined with a precision exceeding
$\mathrm{10^{-11}}$ (about 0.3 $\rm cm \ s^{-1} $).  The integer n is a large number that projects the precision of the atomic clock from the radio regime (the frequency range typical of the atomic clock transitions) onto the optical in the case of the HARPS setup. The design and setup of an LFC dedicated to astronomical applications such as HARPS are described in \citet{2012Natur.485..611W} and \cite{2014SPIE.9147E..1CP}.
In recent years, more studies have been carried out to test the stability and repeatability of the LFCs coupled with HARPS \citep{2020NatAs.tmp...26P, 2020MNRAS.493.3997M}.

Thanks to its spectrum of unresolved, densely spaced, and unblended lines, we are able to use the LFC as a tool to characterize the PSF of HARPS in a large part of its spectral domain and to address various effects that might impact the RV precision.
By 'unresolved' we mean that the intrinsic linewidth of the LFC is negligible compared to the width of the instrumental PSF.
 It has been demonstrated that the HARPS+LFC system reaches a stability of  ~1 $\rm cm \ s^{-1} $  \citep{2020NatAs.tmp...26P} on a short timescale and under stable conditions; however, on a long timescale, and taking into account varying operating conditions, we detect PSF variations that, if not corrected for or calibrated, might reduce the achievable precision. Multiple RV instruments now utilize LFCs, and therefore the material discussed in this paper is relevant for them as well.

In this work, we try to assess the dependence of the PSF shape on various parameters, and we attempt a calibration of the line shift
generated by a varying PSF.
We do not discuss the performances of the LFC here, but we use the LFC spectrum to study the HARPS PSF. We take advantage of the diverse conditions at which data were acquired during the test run to characterize instrumental and detector effects.
A similar characterization using only the lines from a 'classical' ThAr lamp would be less accurate
due to the non-homogenous line intensity and line spacing. In addition, we should note that even having a perfect calibrator is not sufficient for reaching the best precision.
This is possible only if the whole chain of data acquisition and data reduction is carefully optimized, especially for the forthcoming instruments aiming at centimetre per second
RV precision level, such as ESPRESSO \citep{2014AN....335....8P, 2018haex.bookE.157G}, HIRES \citep{2016SPIE.9908E..23M}, EXPRES \citep{2016SPIE.9908E..6TJ}, and NEID \citep{2016SPIE.9908E..7HS}.



Given the advantages of LFCs, we can also study the detector effects that limit the current Doppler measurement precision.
One of the most important detector effects is charge transfer inefficiency (CTI).
During the read-out process, the transfer of charge moving from one pixel to the adjacent pixel is not perfect, leaving a few electrons behind along the clocking path.
This effect is tolerable when a large number of electrons are involved in CCDs .
However, it becomes unacceptable with a low signal level, especially for high-precision RV measurements: CTI affects the instrumental profile and distorts the spectral lines shape, introducing an asymmetry to the spectral lines and inducing systematic errors on the line position that will manifest as a Doppler shift.

In recent years, the CTI effect has been detected and corrected for the SOPHIE spectrograph. The same effect exists in the HARPS detectors, but it is much smaller than in SOPHIE \citep{2000ASSL..252...25C, 2009EAS....37..247B}.
A well-known relation between CTI and the signal level is described in \citet{2006PASP..118.1455G, 2009EAS....37..247B}:
\begin{equation}
 \centering
    CTI(x)=\alpha \cdot I(x)^{-\beta} \cdot exp\left( -0.2 \cdot \left( \frac{B(x)}{I(x)}  \right)^{3} \right)
     \label{e5}
.\end{equation}
Here, $I(x)$ and $B(x)$ are the measured signal and background levels on pixel $x$.
The accuracy of this approach is limited by the precision with which the model parameters can be calibrated, and it is computationally intensive.
Although this approach is perhaps the only way to correct CTI for any generic spectrum, these limitations make this method intractable for HARPS.
Rather, in this work, we simplify the problem by studying PSF variations as a function of signal level and position in the detector using LFC data in both the object and reference fibres.
The LFC lines are distributed
in various positions along the charge transfer path and can be controlled at different flux levels, enabling us to model the CTI effects in the 2D flux-coordinate space.
Consequently, we can use this model, specific to the LFC spectra,  to correct for or mitigate the CTI-related effects in our data set. This is the first time an LFC has been used to accurately study the PSF of a spectrograph and its variations.



This paper is organized as follows. Section \ref{s2} concisely describes the instrumental setup of the HARPS-LFC test system. In Sect. \ref{s3}, we present our line profile analysis on the raw spectra and demonstrate the variation of line profiles as a function of line position and line intensity.
Finally, we provide a discussion and corresponding conclusions in Sect. \ref{s4}.

   \begin{figure}
   \centering
   \includegraphics[width=\columnwidth]{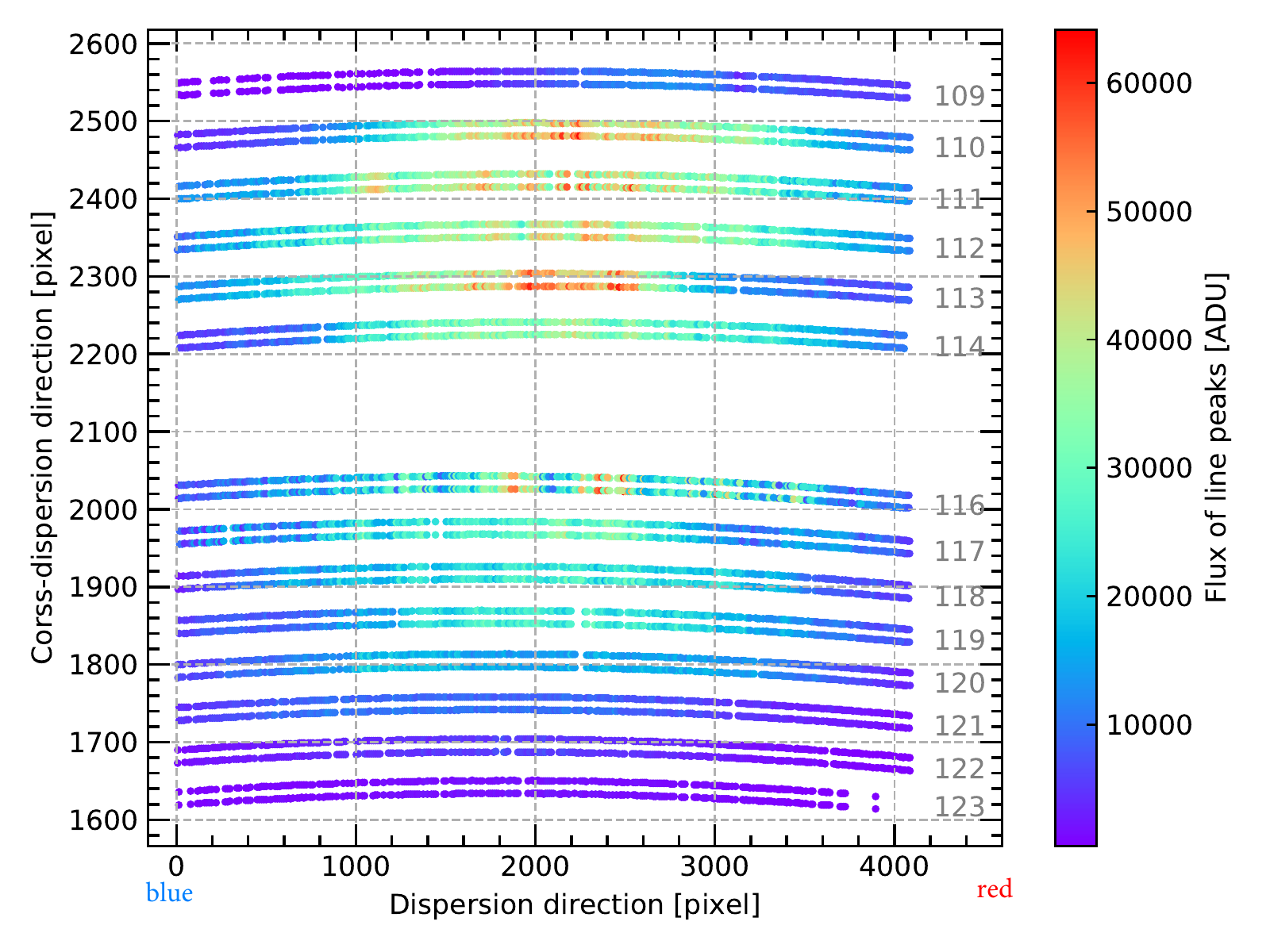}
      \caption{Visualization of the LFC raw spectrum on the CCDs. The peak intensity of the detected LFC lines is displayed via the colour code. Each echelle order's number is marked on the right end of each order curve. The two 2k$\times$4k CCDs are matched together at the gap between order 114 and 116. The serial register in this figure is to the right, at redder wavelengths.}
         \label{Fig2}
   \end{figure}

\section{Instrumental setup and data sample}   \label{s2}

The experimental setup is described in detail in the work of  \citet{2012Natur.485..611W}.
Here we will just briefly state that we used: a mode-locked Yb-doped fibre laser (central wavelength of 1040nm) in conjunction with Fabry-Perot cavities for the suppression of unwanted modes; a second harmonic generation stage; and a photonic crystal fibre for spectral broadening. The output spectra are centred at 530nm and are about 100nm broad, and the delivered mode repetition rate is 18GHz.
The LFC light is delivered to a multi-mode fibre that is stressed and agitated by a small cell phone motor for modal noise mitigation.
Finally, the light is transmitted into the HARPS fibres by mean of a lens doublet, and the recorded spectra show about 4000 lines in each fibre and cover 12 orders in most cases (see Fig.~\ref{Fig2}). The tests performed during the run used a wide variety of possible configurations and signal levels.
The LFC lines are unresolved by HARPS, and therefore the acquired comb spectra are ideal for investigating the variability of the instrumental line profile under various conditions.

The relevant observable for RV determination is the position of the spectral lines; we therefore studied the effects that modify the instrument profile and affect the measurement of the line position.
HARPS uses a second fibre to monitor and correct for instrumental drifts under the assumption that such drifts are the same on the science fibre (Fibre A, which acquires the astronomical target) and the second reference fibre (Fibre B, which acquires calibration light) \citep{1996A&AS..119..373B}.

%
\begin{table} \setlength{\tabcolsep}{3pt} \renewcommand{\arraystretch}{1.3}
\centering
\caption{Exposure series studied in this work. The first series is the 'control' series that was obtained in unperturbed, nominal conditions; it is constituted by 200 spectra. The other series were obtained by changing the flux through the use of the ND filter.  In each of these series, the number of acquisitions is ten. The 'Flux' column shows the average peak flux (not the gross flux) of all the comb lines in each acquisition. The symbol 'B' in series 17-27 indicates that the ND filter was installed on fibre B only.}             
\label{Table1}      
\begin{tabular}[c]{c  m{50pt}  m{50pt} m{50pt}}    
\hline\hline                 
Series No. & Attenuation [dB]  & Exposure time[s] & Flux [ADU] \\  
\hline                        
Control & no filter & 40 & 14527.35 \\
\hline
   01  & 0 & 40  & 13778.52 \\      
   02  & 5 & 40  & 8691.26 \\
   03  & 6 & 40  & 6721.98\\
   04  & 7 & 40  &  5445.81\\
   05  & 8 & 40  & 4605.38 \\
   06  & 10 & 40 & 3163.14   \\
   07  & 11 & 40 & 2713.96   \\
   08  & 12 & 40 & 2203.20   \\
   09  & 13 & 40 & 1810.34  \\
   10  & 13 & 800 & 22840.81    \\
   11  & 14 & 40 & 1491.33   \\
   12  & 16 & 40 & 1065.87   \\
   13  & 18 & 40 & 769.25   \\
   14  & 20 & 40 & 595.21   \\
   15  & 10 & 40 & 3107.71   \\
   16  & 0 & 40 & 22695.24   \\
\hline
17 & 0,B &  40 & 15274.05 \\
18 & 10,B & 40 & 8769.29 \\
19 & 2,B &  40 & 13433.66 \\
20 & 18,B & 40 & 8158.61 \\
21 & 4,B &  40 & 11530.33 \\
22 & 16,B & 40 & 8320.04 \\
23 & 6,B &  40 & 1032.15 \\
24 & 14,B & 40 & 8919.45 \\
25 & 8,B &  40 & 10176.83 \\
26 & 12,B & 40 & 9260.86 \\
27 & 0,B &  40 & 15272.96 \\

\hline                                   
\end{tabular}
\end{table}

A single well-exposed LFC spectrum on HARPS is already sufficient for detecting various features in the line profiles, which would not be detectable with other methods. During our test on the LFC-HARPS system, more than 1700 consecutive calibration spectra were acquired. From the various tests performed, we focus here on: a 'control' sequence that was acquired in unperturbed and nominal conditions; the sequence of spectra obtained with an intervening absorber (neutral density filter) on both fibres (see Table.\ref{Table1}) to change the overall signal level; and a sequence where the absorber was placed at the entrance of Fibre B alone.
We used this setup of neutral density (ND) filters, which covers the full range of linearity on the HARPS detectors, in order to demonstrate how the line intensity affects the line profile.
The conditions at which the data were acquired were chosen to monitor various effects in the whole system and are therefore far from the standard stable operating conditions.
We note that the gain of HARPS detectors is 1.34 $e^{-}$/ADU and the read-out noise is less than 4 ADU in each chip.

The distribution of the LFC lines in a typical spectrum is shown in Fig.~\ref{Fig2}.
The serial register of the detector is located on the right-hand side of the panel. The intensity distribution is largely dependent on the shape of the blaze function and on the output intensity of the LFC.
One pixel on the CCDs has the size of 15$\mu m$, which corresponds to an average Doppler shift of $\approx$ 825$\mathrm{m\,s^{-1}}$.

\section{Characterizing the HARPS PSF} \label{s3}

In this section, we implement a basic moment analysis method to analyze the 2D raw data and study line positions, widths, and symmetries without making assumptions regarding the functional shape of the line profile.

\subsection{Moment analysis}  \label{s3.1}

We studied the instrument PSF via the spectral line's first raw moment (i.e. position),
\begin{equation}
 \mu = \sum_i x_i I(x_i)
      \label{mu}
,\end{equation}
the second central moment (i.e. width),
\begin{equation}
 \sigma^2 = \sum_i (x_i - \mu)^2 I(x_i)
      \label{sigma}
,\end{equation}
and the third normalized central moment (i.e. skewness, to measure the line asymmetry),
\begin{equation}
 \gamma = \sum_i \left(\frac{x_i - \mu}{\sigma}\right)^{3} I(x_i)
      \label{gamma}
.\end{equation}
Here, $i$ is the index and $x_i$ is the $i$th pixel position on the detector. I(x) is the normalized distribution of the the analog digital unit (ADU)
counts.

For each line, we defined a fiducial area around its peak. Rows are parallel to the main dispersion direction. Columns are perpendicular.
We computed the moments for each line using Eqs. \ref{mu}, \ref{sigma}, and \ref{gamma},
after integrating the spectrum along each column.

   \begin{figure}
   \centering
   \includegraphics[width=\columnwidth]{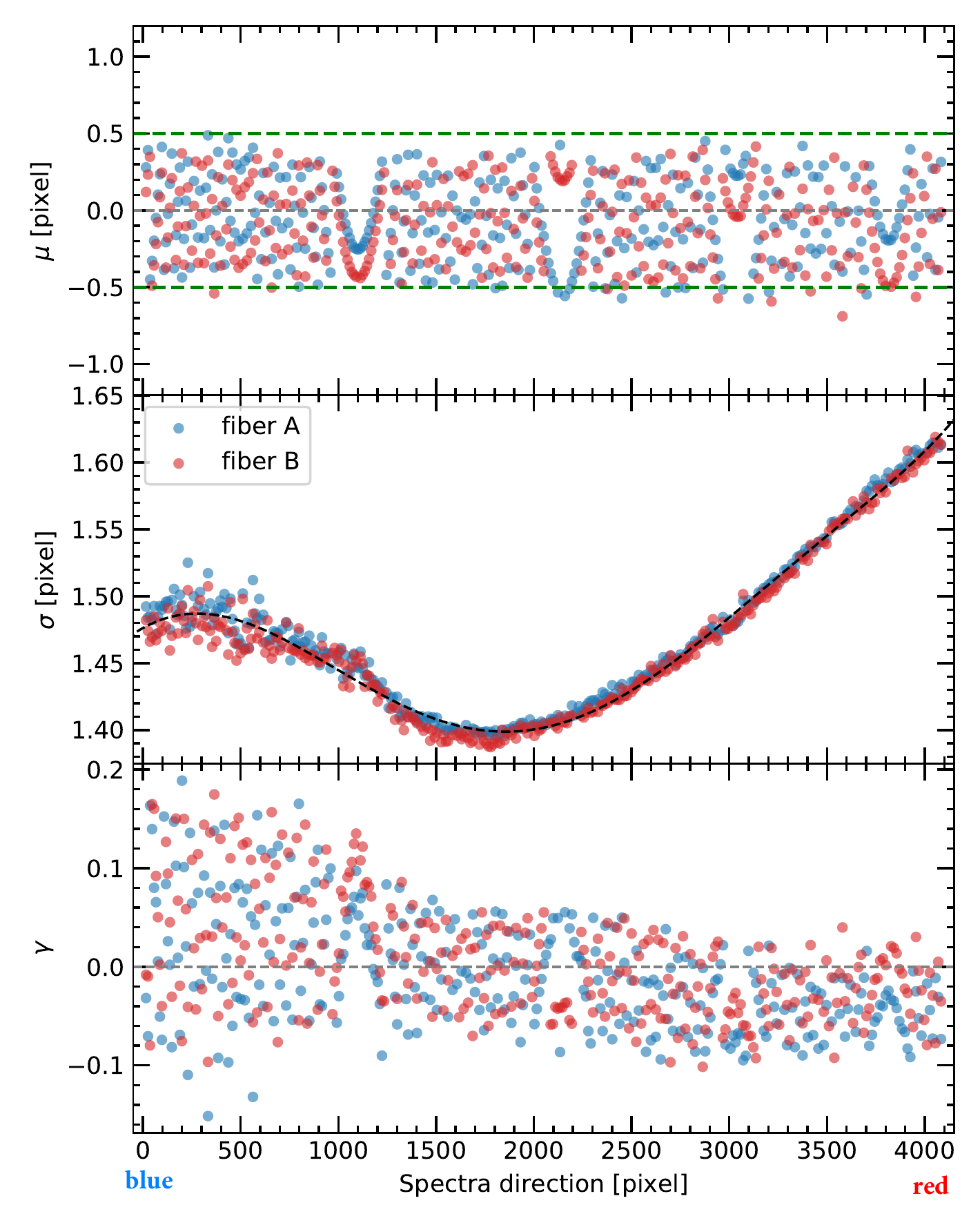}
      \caption{Visualization of moments varying across the detector. The top panel shows the first raw moment ($\mu$) relative to the peak position (in pixels) of the 46th echelle order from a typical acquisition. The middle panel shows the distribution of the square root of the second central moment ($\sigma$). The bottom panel shows the normalized third moment ($\gamma$), or skewness, which defines the asymmetry of the lines. The serial register of the detector is located on the right-hand side, towards the redder wavelength. }
         \label{Fig3}
     \end{figure}

{\it Definition of the area (box).}
The LFC modes separation is about nine to 12 pixels in the HARPS detectors, depending on the wavelength.
For each LFC spectral line measured in the CCD, the peak is defined as the pixel with the highest electron content. An initial 9$\times$9 $pixel^2$ box centred on the peak is defined.
The width of this initial area is based on the estimated typical PSF ($\approx 3.5$ pixels full width at half maximum, $FWHM$, for HARPS).

We increased this area one pixel at a time (in both directions on the detector) as long as the value of $I_{max} - I_{min} $ remained lower than three times the standard deviation of the electron counts within the added area. This process was repeated up to a maximum of 11$\times$11 $pixel^2$.
In a typical spectrum, we find that 98.5\% of the strong lines (ADU
counts at peak > 10000 after background subtraction) are associated with 9$\times$9 $pixel^2$ boxes. Taking into consideration the fact that the faint lines are dominated by noise far from the wings of the PSF, we simply adopted the fixed 9$\times$9 $pixel^2$ box for all the individual comb lines for the moment analysis.

{\it Background subtraction.}
The background was estimated by taking the average of the four pixels with the lowest ADU
counts in the area around the peak, as defined in the previous paragraph, and subtracting the average from the total pixels.

{\it Line selection criteria.}
We selected lines with more than 300 ADU counts after background subtraction at their peak
and used only modes with identified lines in both fibres.

For the 46th order (which is equal to the 113th physical echelle order) of a typical exposure, centred at 542nm, we computed the first raw moment and the second central moment for all lines in both fibres.
The top panel of Fig.~\ref{Fig3} shows the offset between the first raw moment and the peak position of the lines.
A semi-periodic pattern is clearly visible along the main dispersion direction. 
This pattern is due to a quantization error, an alias between the continuous profile of each comb line and the digital representation of the detector. This error has been well reproduced in numerical simulations.

The middle panel of Fig.~\ref{Fig3} shows the second central moment (proportional to the FWHM via the factor $\approx 2.35$).
A third-order polynomial can be fitted to the distribution of the second central moment.
The minimum value is around the pixel position 1720, near the middle of the detector.
The skewness also varies across the CCDs, as seen in the bottom panel of Fig. \ref{Fig3}.

The variations in the shape of the PSF across the detector are the consequences of both optical and detector effects.
Generally, the optical effects may comprise the anamorphism (producing a larger FWHM in the red than in the blue),
the spectrograph image quality, and the uniformity of the light injection (which might vary in time).
However, the dominant detector effect is generally CTI, which is distinguishable from the optical effects due to its dependence on the intensity of the spectrum.

   \begin{figure}
   \centering
   \includegraphics[width=\columnwidth]{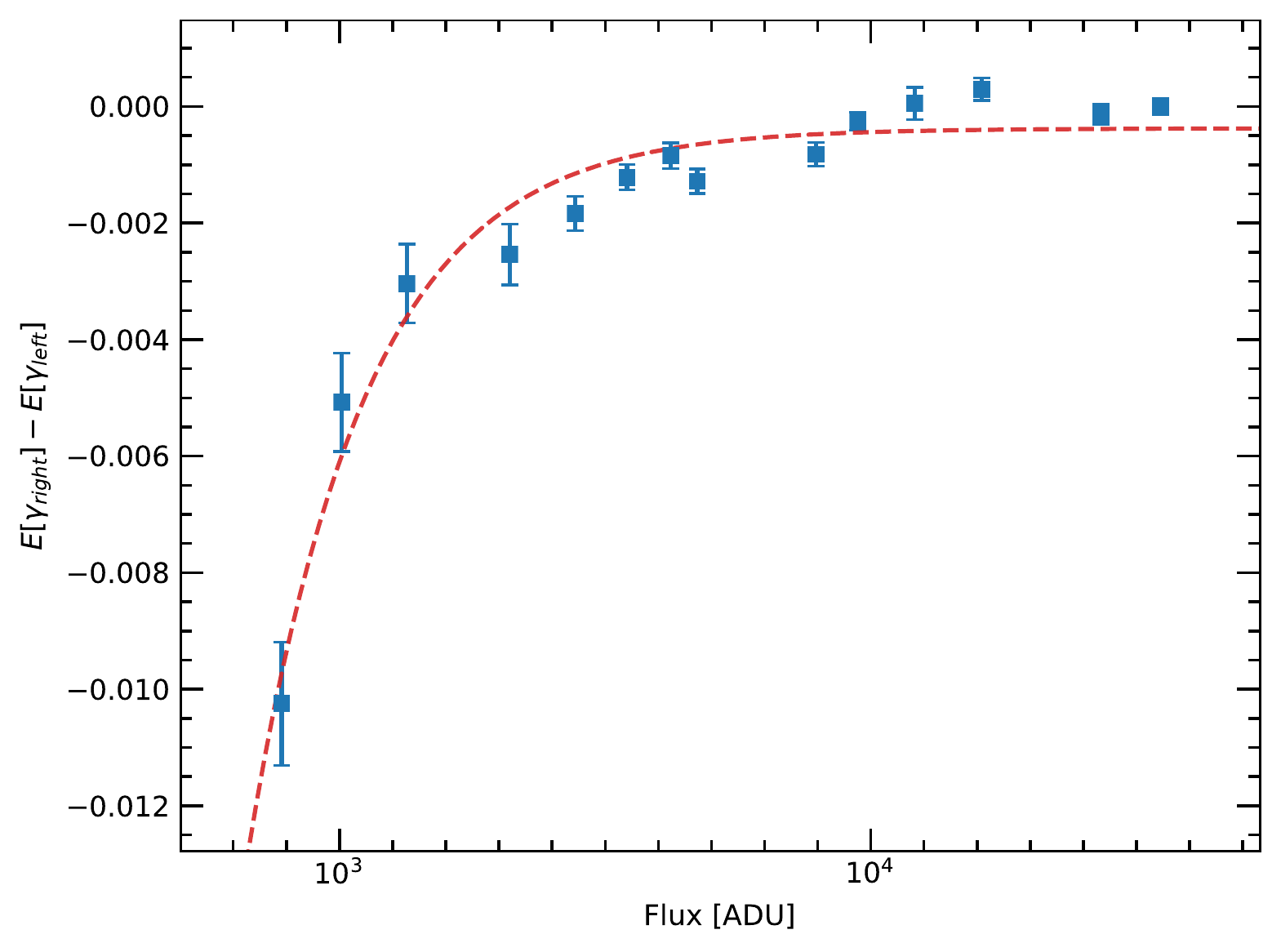}
      \caption{Differential variation (the closer to the right the points are, the closer they are to the serial register
      ) of the normalized third moment, $\gamma$, as a function of flux (averaged peak flux of the comb lines in log scale). The range of the left part of the detector
      goes from the first to the 2048th pixel along the dispersion direction on detectors. The right part of the detector goes from 2048th to the 4096th pixel. Operator $E$ stands for the averaged value for each side on the chip. The dashed red line is an exponential fit to the data points.
      }
   \label{Fig4}
   \end{figure}

 \subsection{PSF dependence on signal intensity} \label{s3.2}

It is important to characterize in detail how the CTI affects the asymmetry of line shapes on HARPS and the estimate of the line positions.
Here, using the LFC lines as a diagnostic tool, we investigated sequences of spectra that were acquired by interposing various ND filters between the output of the LFC and the entrance into the HARPS calibration fibres (Table \ref{Table1}).

The 2D line profile may be distorted by both the horizontal and vertical CTI. In the present analysis, we only focus on the horizontal (spectral direction) CTI. Line profile variations in the vertical direction are less important because they are removed when the spectra are extracted by integrating each order aperture.

 \subsubsection{Skewness} \label{s3.2.1}

To measure the PSF variation across the detector as a function of signal level and position on the CCD, we began dividing the chips into two halves along the main dispersion direction. The right half (pixel $>2048$) is closer to the serial register. Observing the difference in the third moment between the right (closer to the serial register, red part) and the left (far from the serial register, blue part) sides of the orders as a function of signal level, we have a clear indication of CTI.


   \begin{figure}
   \centering
   \includegraphics[width=9cm]{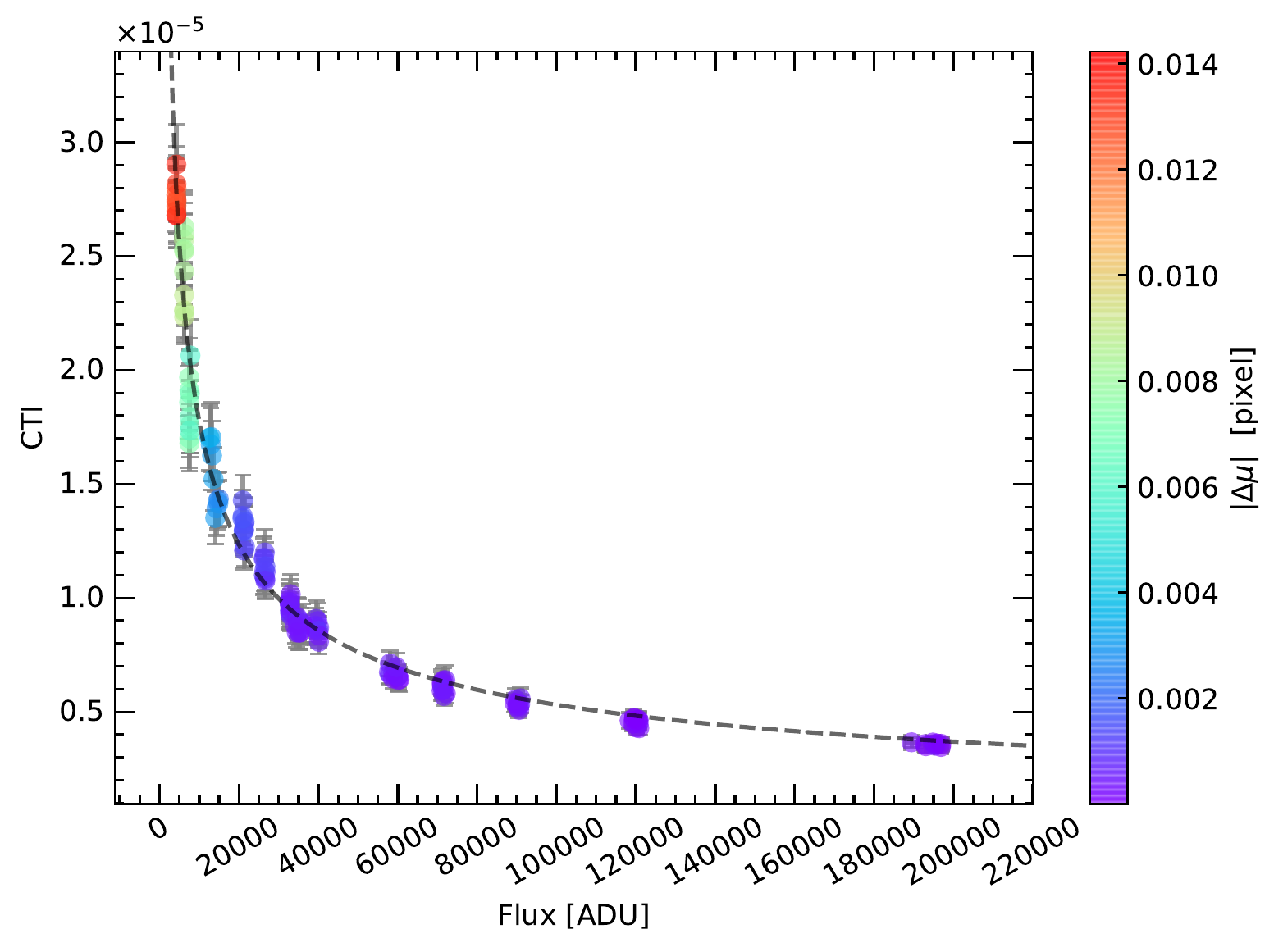}
      \caption{
      Correlation between the CTI indicator and the integrated line flux.  The shift  $\Delta \mu$ of the line's centre of gravity in reference to the first spectrum of series 01 (see Table.\ref{Table1}) is shown via the colour scale.}
         \label{Fig5}
   \end{figure}

Figure \ref{Fig4} shows that the line shape varies differently in the two halves of the detector as the signal intensity level changes. All available LFC lines were used in Fig. \ref{Fig4}. The relationship derived from the fitted curve (dashed line in Fig. \ref{Fig4}) is
$\Delta E[\gamma] = -4.25 \times 10^{3} \cdot Flux^{-1.95} - 3.75 \times 10^{-4}$,
where $\Delta E[\gamma]$ is the difference in the averaged third moment between the right and left parts of the detector;  here, the flux refers to the averaged peak of comb lines.

In presence of CTI effects -- which are expected to be stronger at lower signal levels, as shown in Eq. \ref{e5} -- we expect the lines far from the serial register to develop a longer tail at lower signal levels, while lines closer to the serial register, which suffer from fewer charge transfers, will be less affected. This tail would point away from the position of the serial register. Given the HARPS spectral format, the implication of this tail is an effective blue-shift of the lines for a higher CTI. In practice, a spectrum with a lower S/N (i.e. higher CTI) will be blue-shifted with respect to a spectrum with a higher S/N. At high signal levels, the CTI effect is minimal (Eq. \ref{e5}), and indeed our indicator in Fig. \ref{Fig4} approaches zero.


   \begin{figure*}
   \centering
   \includegraphics[width=\textwidth]{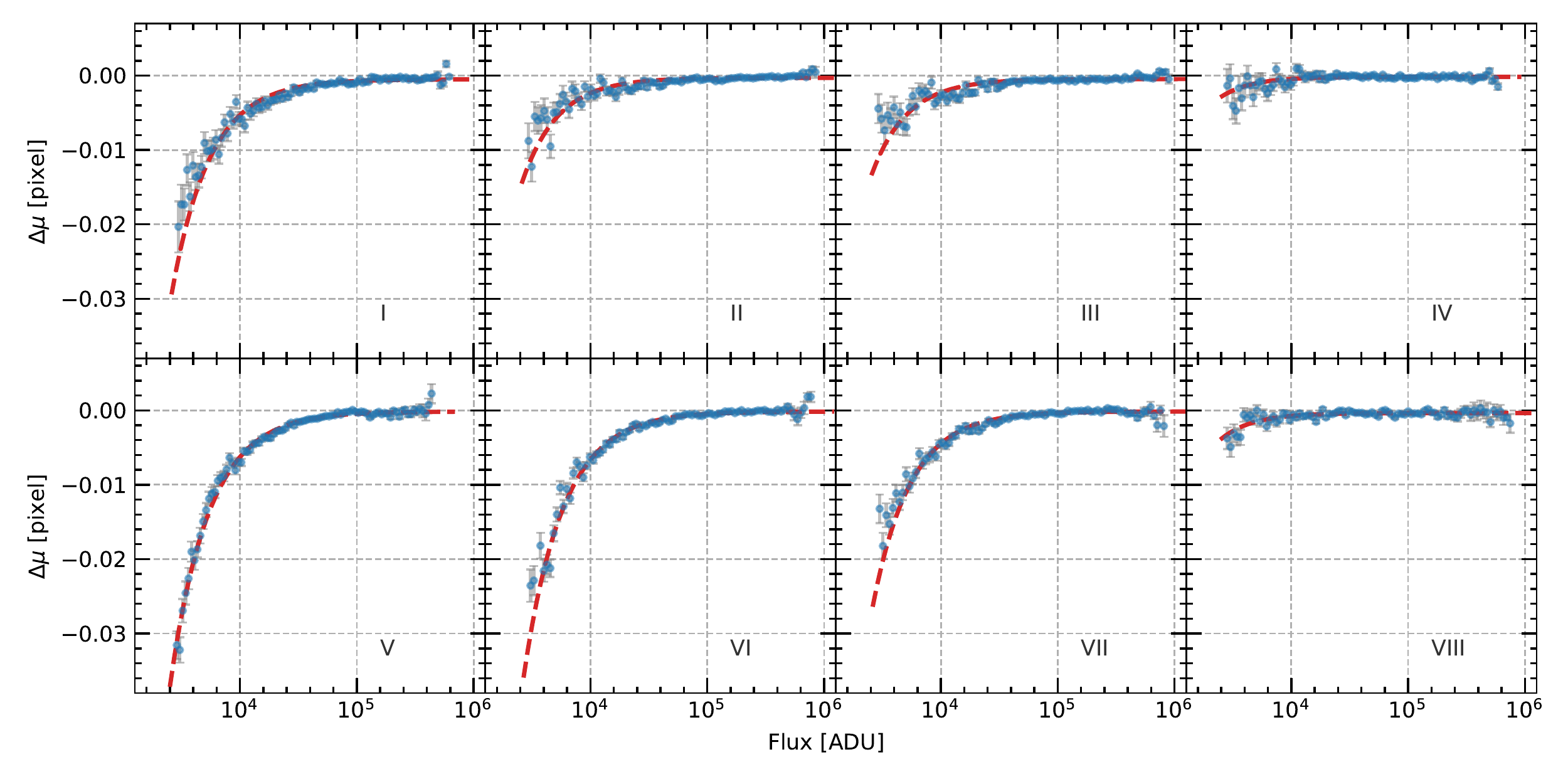}
      \caption{Correlation between line centroid and integrated line signal intensities (in log scale) in different areas of the CCDs. Each CCD was divided into four stripes that are parallel to the cross-dispersion direction, and each stripe covers an interval of 1024 pixels along the main dispersion direction. All LFC lines from the 160 spectra were used to produce this plot. Lines are grouped into 100 flux bins in each subplot (stripe). Exponential functions are used to fit the line shifts as a function of the measured line flux. The fit result is shown with the dashed red line. The relative drift $\Delta \mu$ is measured in pixel space, compared with the reference acquisition, i.e. the acquisition with the highest averaged counts (or lowest attenuation).
      }
         \label{Fig6}
   \end{figure*}

 \subsubsection{CTI} \label{s3.2.2}

The distance between the position of the line peak and the first moment (or centre of gravity) is also an indication of the line asymmetry; this is also the case for the third moment.
It should be noted that we do not try to measure the real experimental CTI values here: We only use the characteristics and distribution of comb lines to reveal and enhance the line intensity dependence of the CTI effect on the current CCD.
%
\begin{table*} \setlength{\tabcolsep}{8pt} \renewcommand{\arraystretch}{1.5}
\centering
\caption{Coefficients used in Fig. \ref{Fig6} to calculate curves for fitting each part of the detector.}             
\label{Table2}      
\begin{tabular}[c]{c | c   c   c }    
\hline\hline                 
Sections & $ a $ &  $ b $ & $\chi^2_{red}$ \\  
\hline                        
   I  & $(-4.98\pm0.45) \times 10^{-4}$  & $0.327 \pm 1.2\times 10^{-3}$ &  2.64 \\
   II  & $(-2.98 \pm 0.30 )\times 10^{-4}$   & $ 0.306 \pm 2.1\times 10^{-3}$ &  2.94 \\
   III  & $(-4.61 \pm 0.29 )\times 10^{-4}$   & $ 0.303 \pm 2.2\times 10^{-3}$ &  3.25 \\
   IV  & $(-1.71 \pm 0.25)\times 10^{-4}$   &  $0.265 \pm 6.5\times 10^{-3}$ &  1.59 \\
   V  & $(-1.66 \pm 0.43 )\times 10^{-4}$   & $ 0.333 \pm 5.2\times 10^{-4}$ &  1.71 \\
   VI  & $(-1.53 \pm 0.38)\times 10^{-4}$  & $ 0.335 \pm 6.8\times 10^{-4}$ &  2.87 \\
   VII  & $(-1.55 \pm 0.37)\times 10^{-4}$  & $ 0.324 \pm 9.2\times 10^{-4}$ &  3.01 \\
   VIII  & $(-3.73 \pm 0.41)\times 10^{-4}$  & $ 0.270 \pm 3.9\times 10^{-3}$ &  1.79 \\
\hline                                   
\end{tabular}
\end{table*}

We began with a model that initially has an unresolved emission line that is fully contained within a pixel, and we shifted it pixel by pixel, taking the CTI\ effect into account.
Once the charge $I$ is shifted from pixel $X$ to pixel $X-1$,
the charge $I(1-CTI)$ will be in pixel $X-1$ and the charge $I\times CTI$ will be left in pixel $X$, and so on.
Finally, we calculated that the error we suffer by estimating the line position with a 'centre of gravity' estimator (very similar to a Gaussian estimator) is approximately:
\begin{equation}
 \centering
        \Delta X(pixels) \approx X\times CTI
     \label{eq_xcti}
.\end{equation}

Therefore, we can give a prescription for correction once we have the value of the CTI, which we know is signal-dependent.
We defined a CTI indicator estimated on the LFC line $i$ (which is proportional to the measured CTI value) as follows:
\begin{equation}
\begin{aligned}
  &  CTI \approx \frac{\Delta \mu}{p} \\
  &  \Delta \mu = (\mu - p)_i - (\mu - p)_{i, 0}  \\
     \label{e6}
\end{aligned}
.\end{equation}
Here, $\mu$ refers to the line's centroid as defined in Eq. \ref{mu} and $p$ is the line's peak position in pixels.
The ($\mu-p)_i$ is the difference between the computed $\mu$ and the peak $p$ on the $i$th line.
The $(\mu-p)_{i,I_{0}}$ is the difference ($\mu-p$) measured for the same $i$th line in the reference frame of the series (i.e. the first spectrum of the sequence with a 0 dB filter). 
It is noted that the definition described here is a CTI indicator, which is not exactly the CTI calculated with Eq. \ref{e5}.

   \begin{figure*}
   \centering
   \includegraphics[width=\textwidth]{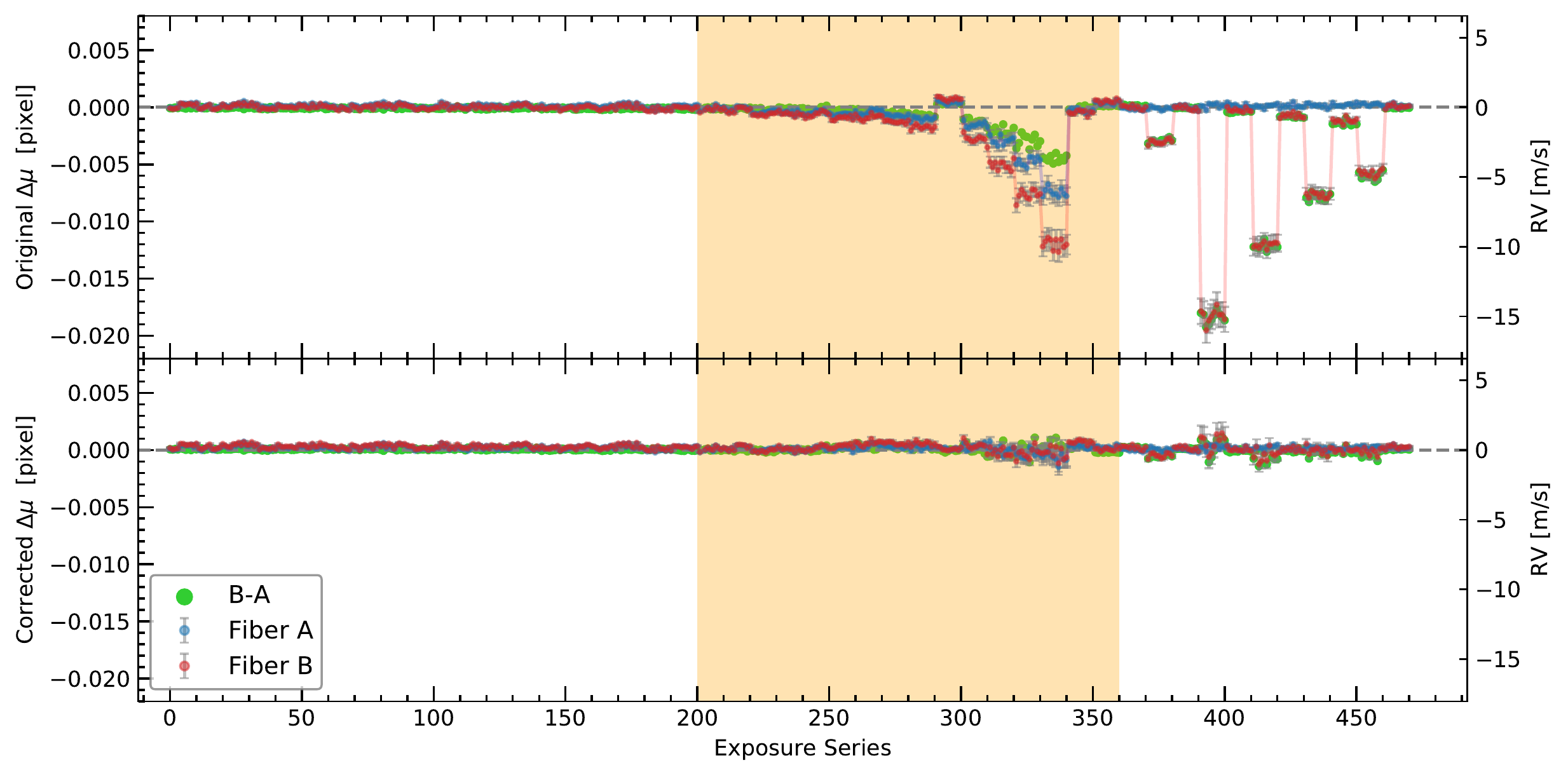}

   \caption{Line shifts of the three sequences studied here: an unperturbed sequence of 200 exposures (left), the sequence of 160 exposures used to calibrate the signal intensity dependence of the CTI (shaded area), and a further sequence where only the absorber at the entrance of fibre B was varied (right). The line positions are compared to a reference obtained with a 0 dB filter. The top panel shows the original data, and the bottom panel shows the data after the correction.}

         \label{Fig7}
   \end{figure*}

Figure \ref{Fig5} illustrates, for all 160 spectra (one point per spectrum): the averages over one spectrum weighted by each line's intensity; the CTI indicator as a function of the integrated line flux; and the centroid shift $\Delta \mu$.
When we fit the curve according to Eq. \ref{e5}, the parameters we derive are $\alpha$ = (2.2 $\pm 0.1)\times 10^{-3}$, $\beta $ = 0.521 $\pm$ 0.007.
In the fit we used the average value of the background B(x) over the whole chip, a typical value of 0.34 ADU per pixel. The reduced $\chi^2$ value derived from this fit is three; the main deviation from the fit is at high fluxes, a region that is poorly sampled. The fact that we can fit the CTI curve nicely to our CTI indicator is further evidence of the validity of this indicator, which is proportional to the real CTI.


To estimate the impact on the RV measurement, we used the average RV content of one HARPS pixel, 825 $m \ s^{-1}$, and assumed a 'good' CTI value of $ 1 \times 10^{-6}$.
Following Eq. \ref{eq_xcti}, this causes an average offset across one order of $\sim 1.8m \ s^{-1}$. If we think that the CTI can vary by up to 50\% due to line intensity variation (especially at a low signal level) -- which is an assumption we have generally made when observing absorption lines of astronomical objects -- we see that this introduces an RV variation of the order of magnitude of $\approx 1m \ s^{-1}$.
This is among the reasons why it is important to always acquire data at about the same S/N, so as to avoid RV variations caused by changing CTI. However, while this strategy might be sufficient for the 1 $m \ s^{-1}$ RV measurement instruments, it is clearly inadequate when aiming at a precision of a few centimetre per second.



 \subsection{Centroid correction } \label{s3.3}

To investigate the dependence of the measured line position (via first moment analysis) with signal intensity, we divided each CCD chip into four sections along the main dispersion direction. The sections each cover 1024 pixels. In Fig. \ref{Fig6}, the top four panels (I, II, III, IV) show the red chip, whereas the four bottom panels (V, VI, VII, VIII) correspond to the blue chip. The serial register is situated close to sections IV and VIII, which are the least affected by signal intensity variations.

From Fig. \ref{Fig6}, we can see that the signal intensity dependence gradually varies along the main dispersion direction towards the serial register position on the right edge.
The red curves in Fig. \ref{Fig6} are exponential fitting functions for each part (from I to VIII) with the form of:
\begin{equation}
 \centering
    \Delta \mu = a_i - A\cdot e^{-\left(\tfrac{I}{b_i}\right)},
     \label{e7}
\end{equation}
where $i=I, II, III, ... VIII$, $A = 10^3$, and $I$ is the flux (X-axis in Fig. \ref{Fig6}).
The coefficients $a$ and $b$ are listed in Table \ref{Table2}.
We note that the range we explore with our data points does not allow us to fit the amplitude $A$ of the exponential trend as a free parameter.
It therefore seems  reasonable to fix the amplitude to a constant value that allows for a reliable fit.
The fit is not optimum, but this is the best we can do with the data we have.

We then used these eight functions to correct for the line profile distortions with both the position and line intensity dependence. The correction was applied to each individual comb line in the test exposures one by one. For example, we determined a comb line $i$ with flux $I$  on the top chip and a 2500 pixel position in the main dispersion direction to be in area III, and we automatically applied function III  to calculate the $\Delta \mu$. Then we took this $\Delta \mu$ as the offset to this line to be corrected for.
 The exposure samples we chose are described in Table \ref{Table1}, where we show the results of the 16 groups of spectra (each group has ten exposures) that we tested by changing the flux levels. The flux was changed via the interposition of an absorber. Figure \ref{Fig7} shows the comparison before and after the line distortion correction. The top panel displays the original drifts of the examined
 samples. The corrected results are displayed in the same Y-axis frame scale in the bottom panel.
%

\begin{center}
\begin{table*} \setlength{\tabcolsep}{3pt} \renewcommand{\arraystretch}{1.4}
\centering                
\caption{RMS of the line positions from Fig. \ref{Fig7}. 'A' represents fibre A, and 'B' represents fibre B. The subscript 'o' stands for 'original', while 'c' stands for 'corrected'. The RMS is presented in two units (pixel and $cm \ s ^{-1}$). These values are listed in different lines in each concerned box.}        
\label{Table3}      
\begin{tabular}[c]{c  | c  c  c | c  c  c | c}        
\hline\hline                 
Number & $rms(A)_{o}$  & $rms(B)_{o}$  & $rms(A-B)_{o}$  & $rms(A)_{c}$  & $rms(B)_{c}$  & $rms(A-B)_{c}$  &    \ \ \ \  \\    
of spectra & [pixel] & [pixel] & [pixel] & [pixel] & [pixel] & [pixel] &    \ \ \ \ \ Notes  \\    
             & [$\rm cm \ s^{-1} $] & [$\rm cm \ s^{-1} $] & [$\rm cm \ s^{-1} $] & [$\rm cm \ s^{-1} $] & [$\rm cm \ s^{-1} $] & [$\rm cm \ s^{-1} $] &    \ \ \ \   \ \\    
\hline                        
  200   &  1.97$\times 10^{-4}$  &    1.73$\times 10^{-4}$  &  1.06$\times 10^{-4}$   &  2.87$\times 10^{-4}$  &    3.39$\times 10^{-4}$  &  8.89$\times 10^{-5}$  &  Quiet Series \\
            &   16.3              &    14.3              &  8.7                   &  23.7            &    28.0             &  7.3                &   A-B at photon noise\\
\hline
160   &  2.42$\times 10^{-3}$  &    3.92$\times 10^{-3}$  &  1.53$\times 10^{-3}$   &  3.42$\times 10^{-4}$  &    4.33$\times 10^{-4}$  &  3.11$\times 10^{-4}$  &  Standard series, as in Fig.\ref{Fig7}, yellow band \\
                       &   199.7            &    323.4            &  126.2               &  28.2 &    35.7              &  25.7               &   A-B at photon noise in the corrected data\\

\hline
110   &  2.05$\times 10^{-4}$  &    7.27$\times 10^{-3}$  &  7.35$\times 10^{-3}$   &  2.22$\times 10^{-4}$  &    4.38$\times 10^{-4}$  &  4.41$\times 10^{-4}$  &  ND filters only on fibre B \\
         &   16.9           &    599.8           &  606.4               &  18.3             &    36.1              &  36.4                &   B-A at photon noise in the corrected data \\
\hline
\end{tabular}
\end{table*}
\end{center}

The improvement in the line stability, after the correction, is evident from Fig. \ref{Fig7}.
The main result is summarized in Table \ref{Table3}.
It demonstrates the shifts (in pixels and in RV) of the lines of the spectra in the sequences listed in Table \ref{Table1}, for a total of 470 spectra. The yellow shaded area in the figure indicates the spectra used for the calibration of the CTI correction. The bottom panel of the figure shows the corrected data. The numerical results are shown in Table \ref{Table3}.

After the correction, the shift difference between Fibres A and B is consistent with the photon noise.
We notice that Serial No. 10 in Table\ref{Table1}, with an absorption of about a factor of 20 ($\sim$13dB) and an exposure time longer by a factor of 20, does not show any systematic line shifts in either the uncorrected or corrected analyses.
This shows that the line shift is solely dependent on the line flux.
For comparison, a series with 200 spectra that were acquired in stable conditions is presented (the control series). Also in this case (as expected), the correction algorithm gives an RMS of the A-B shift difference at the photon noise level.
A final comparison is made with the series in which an absorber was placed at the entrance of the B fibre only (series 17-27 in Table \ref{Table1}). Also in this case, we are at the photon noise level after the correction.

The current moment analysis results indicate that the shape of the PSF is generally asymmetric on HARPS and that it varies with
the position in the detector and with the intensity level of the lines. Optical, injection, and detector effects all contribute to the overall shape of the PSF, and it is difficult to disentangle their different effects. Nevertheless, we have clearly identified the CTI effect in our data set with the moment analysis method, which can  be modelled and effectively corrected.
The main output of these tests can be summarized as: i) The RMS of the differential drift (drift Fibre A minus drift Fibre B) of the control series was at photon noise before the correction and it is still at photon noise after the correction.
ii) The RMS of the differential drift of the calibration sequences was at 126.2$ cm \ s^{-1}$ before the correction, and it is at photon noise after the correction. iii) The RMS of the differential drift of another set of sequences, not used for the calibration and done when the signal intensity was varied in a different fashion than in the calibration sequence, is at photon noise after the correction.

We conclude that the correction we extracted from our data is capable of providing, at photon noise, sequences of highly varying signal strengths, without detrimental effects on sequences that were already at photon noise: That is, the correction is effective and does not deteriorate the data, at least not the data sample we used to test it.

\section{Summary and discussion} \label{s4}



Based on the significant advantages of the LFC compared to the traditional wavelength calibration sources, the accurate, unresolved, and densely distributed comb lines can be used as a tool to measure and characterize the instrumental effects induced by the telescope, the spectrograph, and the detector systems.
When aiming for an RV measurement precision close to the centimetre per second
level, these effects must be accounted for and, when possible, corrected.


We studied the HARPS line profile  and investigated its variations as a function of line intensity and line position in the focal plane of the spectrograph.
Using moment analysis on the raw 2D data, we demonstrate that the magnitude of asymmetry in the PSF is dependent on the signal intensity and line position.
We find that the LFC lines are more blue-shifted at lower flux levels.
We trace this back to detector effects that can be modelled as CTI.
Indeed, the distortion of LFC line profiles varies with the signal level and with the distance to the serial register, as expected in the case of CTI (See Figs. \ref{Fig4} and \ref{Fig5}).

We show that the line shifts can be parameterized and corrected for, reducing the shifts from metre per second
down to the photon noise of a few centimetre per second.
However, our parametrization is specific to the LFC emission lines.
In principle, line shifts in stellar absorption spectra would scale differently with different intensity levels and positions in the detector.
Considering that the spectral lines of solar-type stars have a contrast of the order of 50\%, and that the typical linewidths are of a few kilometre per second,
we expect the impact of CTI to be comparably low on these wider spectral lines.
Moreover, in standard monitoring observations, for example in a planet search programme, the spectra on the same target are acquired with comparable intensity levels (S/N), that is, the CTI will not differ much among the different spectra.


The understanding of these sources of errors is critical when seeking the high RV precision ($\sim$ centimetre per second)
necessary for the detection of Earth-twins in the habitable zone around solar-type stars, the variation of the fundamental constants, and the measurement of the accelerated expansion of the Universe, which will be the main science cases for the the Extremely Large Telescope (ELT)
\citep{2013arXiv1310.3163M, 2015GReGr..47.1843M, 2016SPIE.9908E..23M} in the coming future.

\begin{acknowledgements}
We would like to thank the ESO LFC team, Max-Planck-Institut f\"ur Quantenoptik and Menlo Systems GmbH for providing the data.
All the authors are grateful for the staff at La Silla Observatory for their huge support during the measurement campaigns.
We sincerely thank the anonymous reviewer for the constructive comments and suggestions,
which greatly improved the paper.
We are also grateful to Rachel Baier, Sarah Bird, Frank Grupp, Xiaojun Jiang, Yujuan Liu, Liang Wang and Huijuan Wang for their helpful comments.
This work is supported by the National Natural Science Foundation of China(NSFC) under grant No. 11703052, 11988101 and 11890694.
This study is also performed by the National Key R\&D Program of China No. 2019YFA0405102 and  2019YFA0405502.
The research activity of the Observational Astronomy Board of the Federal University of Rio Grande do Norte (UFRN) is supported by continuing grants from the Brazilian agencies CNPq and FAPERN. JRM, BLCM and ICL also acknowledge continuing financial support from INCT INEspa\c{c}o/CNPq/MCT. ICL acknowledges a Post-Doctoral fellowship at the European Southern Observatory (ESO) supported by the CNPq Brazilian agency (Science Without Borders program, Grant No. 207393/2014-1). This study was financed in part by the Coordena\c{c}\~{a}o de AperfeiÃ§oamento de Pessoal de N\'{i}vel Superior - Brasil (CAPES) - Finance Code 001.
L.Pasquini acknowledges a distinguished visitor PVE/CNPq appointment at the DFTE/UFRN and thanks the DFTE/UFRN for hospitality.
JIGH acknowledges financial support from the Spanish Ministry of Economy and Competitiveness (MINECO) under the 2013 Ram\'{o}n y Cajal program MINECO RYC-2013-14875, and, JIGH, RRL and ASM also acknowledge financial support from the Spanish ministry project MINECO AYA2014-56359-P.
We also thank and acknowledge the support from ESO-NAOC student-ship.

\end{acknowledgements}

\


\bibliographystyle{aa}    
\bibliography{bib}    

\end{document}